\documentstyle[preprint,aps]{revtex}

\begin{document}
\input{epsf.sty}
\draft
\title{
Analytical results for Scaling Properties of the Spectrum of
the Fibonacci Chain}
\author{ Fr\'ed\'eric Pi\'echon$^{a,b}$, Mourad Benakli$^a$ and Anuradha
Jagannathan$^a$}
\address{$^a$Laboratoire de Physique des
Solides, Universit\'{e} Paris--Sud, 91405 Orsay, France}
\address{$^b$Laboratoire de Physique Quantique, Universit\'{e} Paul Sabatier,
31062 Toulouse, France}
%\twocolumn[
\date{February 1995}
\maketitle
%\widetext
%\vspace*{-1.truecm}
%\hspace*{-1.truecm}
\begin{abstract}
%\begin{center}
\parbox{14.cm}{We solve the approximate renormalisation group found by Qiu Niu
and Franco Nori\cite{niunori} for a quasiperiodic tight-binding hamiltonian on
the Fibonacci chain. This enables us to characterize analytically the spectral
properties of this model.}
%\end{center}
\end{abstract}
\hspace*{1.truecm}
\pacs{PACS numbers: 05.45+b 61.44+p 71.25c}
%]
%\draft
%\Date{1/95}
%\debugon
\narrowtext
%\vspace{-1.2truecm}{
In the past few years, many models \cite{review}have shown that a particle
moving on a infinite chain and subjected to a quasiperiodic (QC) or
incommensurate (IC) modulation can exhibit critical localization properties.
Depending on the QC (IC) modulation strength, in contrast to the case of a
disordered modulation, the particle wavefunction is not necessarily strongly
localized (Insulating); it can also be extended (Metallic) as in a
translationally invariant system or in a critical localization regime at the
Metal--Insulator transition. Typical models where these peculiar
localization properties occur are Harper like models (IC) or tight--binding
hamiltonians associated with a quasiperiodic sequence such as the Fibonacci
sequence (QC). In general, these models have two important parameters: an
irrational $\omega $ which is responsible for the absence of periodicity,
and a parameter $K$ which determines the strength of the (QC) or (IC)
modulation. The usual procedure to characterize the localization properties
of these hamiltonians $H_{\omega ,K}$ is to study the scaling of the
spectral properties of a sequence of periodic hamiltonians $H_{\omega _n,K}$
which converge to $H_{\omega ,K}$ as $n$ goes to infinity. If the widths of
the $q_n$ individual bands that compose the spectrum of $H_{\omega _n,K}$,
decrease exponentially with the period $(q_n)$ of $H_{\omega _n,K}$ then,
$H_{\omega ,K}$ is in a insulating regime and has a {\em Pure Point}
spectrum. In the case when the widths decrease only inversely proportionally
to $q_n$, then $H_{\omega ,K}$ is in the metallic regime and has a {\em
Absolutely Continuous} (AC) spectrum. In contrast with these two cases, the
critical localization regime of $H_{\omega ,K}$ is expected to have a {\em
Singular Continuous} (SC) spectrum with multifractal (MF) properties in the
spectral measure. In fact, in this latter case, the scaling found by many
numerical simulations \cite{review} is that the bandwidths decrease like $
\sim q_n^{-1/\alpha }$ where the exponent $\alpha (<1)$ varies from band to
band so that there are typically $\sim q_n^{+g(\alpha )}$ bands with the
same exponent. Due to these peculiarities, the critical regime has been
extensively studied by both numerical simulations and analytic techniques%
\cite{review}. However, as yet, a analytical quantitative determination of
the exponents ,$\alpha $ and $g(\alpha )$, has not been achievable for even
one irrational $\omega $.

The purpose of this paper is to show a QC tight--binding model for
which we are able to analytically
characterize all the spectral properties. Our work starts from the approximate
renormalisation group (RG) found by Niu
and Nori\cite{niunori} for a QC model on the Fibonacci chain. By
reformulating and solving the RG of Niu and Nori, we derive constructive
and transparent recurrence schemes for both the energy levels and the
bandwidths. From these two schemes we deduce new recurrence relations for
the spectral measure, the large time average return probability of particle
defined in \cite{geisel}, the spectrum Lebesgue measure, the MF partition
function \cite{halsey} and the bandwidth distribution. For most of these
relations, a natural fixed point solution is a power law, either in size or
time. By comparing with the fixed point equation of the MF partition function,
it appears that the exponents associated to these power laws, are
related to a subset of the $anomalous$ dimensions which characterize the MF
properties of the spectral measure. A direct calculation of the function $
g(\alpha )\ vs\ \alpha $ confirms these MF properties. To complete the
analysis of the spectral properties, we also study the gap properties. We
find that there are two types of gaps : $transient$ and $stable$. For the
first type, their properties are like those of the bandwidths. In contrast,
for the $stable$ gaps, the distribution of their width ($g$) is a $stable$
power law $P(g)\sim g^{-(1+D_F)}$ where $D_F$ is the Hausdorff dimension of
the spectrum measure.\\ These results complete and correct a previous MF
analysis of the spectral measure of this model\cite{zheng}; our work also
unifies many partial results obtained by other methods, for both this model
\cite{kohmoto1}\cite{kohmoto2}\cite{kohmoto3} and the Harper model\cite
{bellstinchcombe}.

We consider the tight--binding hamiltonian $H_n$ defined on approximant of
period $F_n$ of the Fibonacci chain by the following equation:
\begin{equation}
\label{hamilt}H_n=\sum_{i=1}^{F_n} V_ic_{i}^{\dagger}c_{i}+
t_{i,i+1}c^{\dagger}_{i}c_{i+1} + t_{i-1,i}c^{\dagger}_{i}c_{i-1}
\end{equation}
The on site potential $V_i$ is taken to be uniform ($V_i=V$). In contrast,
the hopping amplitude $t_{i,i+1}$ from site $i$ to site $i+1$ is given by $
t_{i,i+1}=t_w(1-\chi(\omega_n i))+t_s\chi(\omega_n i)$, where $\omega_n\!=\!
\frac{F_{n-1}}{F_n}$ tends to the golden mean $\omega\!=\!\frac{\sqrt{5}-1}{2
}$ ($F_{n+1}\!=\!F_n+\!F_{n-1}$ and $F_n\simeq\omega^n$). The characteristic
function $\chi(\omega i)$ takes the value $0$ or $1$ according to the
Fibonacci sequence and correspondingly, the bond $t_{i,i+1}$ will take the
value $
t_w $ or $t_s$. For finite $n$ the density of weak bonds $(t_w)$ is $\omega_n
$ and tends to $\omega$ in the quasiperiodic limit. For the strong bonds $
(t_s) $ the density is $\omega_{n}^{2}=\frac{F_{n-2}}{F_n}$ and tends to $
\omega^2$. The periodicity of the hamiltonian $H_n(V,t_s,t_w)$ allows us to
define Bloch boundary conditions of the form $c_{j+F_n}=e^{ik}c_j$. For a
fixed $k$, the energy spectrum of $H_n(V,t_s,t_w)$, which we define as $
W_n(V,t_w,t_s)$, consists of $F_n$ levels $E^{i}_{n}(k)$ ($i=1,...,F_n$ and $
E^{i}_{n}\le E^{i+1}_{n}$ by convention). Varying $k$ from $0$ to $\pi$
allows the association of an energy band of width $
\Delta^{i}_{n}=|E^{i}_{n}(\pi)-E^{i}_{n}(0)|$ to each of these levels.

Using a perturbative approach, Niu and Nori have shown that in the strong
modulation regime ($t_w/t_s\ll 1$) the spectrum $W_n(0,t_w,t_s)$ is the
union of three sub--spectra $W_{n-2}(V^{+},t_s^{+},t_w^{+})$, $
W_{n-3}(V^0,t_s^0,t_w^0)$ and $W_{n-2}(V^{-},t_s^{-},t_w^{-})$, which
correspond to three sub--hamiltonians with periods $F_{n-2}$, $F_{n-3}$,
$F_{n-2}$ and
renormalized parameters $V^{\pm ,0}$, $t_s^{\pm ,0}$,$t_w^{\pm ,0}$
respectively. A representation of this perturbative RG, with the explicit
value of the renormalized parameters, is schematically given by the
following relation (\ref{niunori}).
\begin{equation}
\label{niunori}W_n(\displaystyle{0,t_s,t_w})\longrightarrow \left|
\begin{array}{l}
W_{n-2}(t_s,
\frac{\displaystyle{+t_w}}{\displaystyle{2}},\frac{\displaystyle{+t_w^2}}{%
\displaystyle{2t_s}}) \\ W_{n-3}(0,
\frac{\displaystyle{-t_w^2}}{\displaystyle{t_s}},\frac{\displaystyle{t_w^3}}{%
\displaystyle{t_s^2}}) \\ W_{n-2}(-t_s,\frac{\displaystyle{-t_w}}{%
\displaystyle{2}},\frac{\displaystyle{-t_w^2}}{\displaystyle{2t_s}})
\end{array}
\right.
\end{equation}
In principle, the scheme (\ref{niunori}) simplifies the problem, since it
relates the spectral properties of a hamiltonian of period $F_n$ to those
of three sub--hamiltonians of smaller period. However, it is clear that upon
$l$ iterations of (\ref{niunori}), the difficulty which is initially due to
the large period $F_n$, is replaced by the problem of an increasing number $
(3^l)$ of different hamiltonians to be solved. As we now describe, there is
some properties of both the hamiltonian and the RG (\ref{niunori}) allow us to
overrule this difficulty. Firstly, it is clear that the spectrum of $
H_n(0,t_s,t_w)$ is independent of the sign of $t_s$ and $t_w$, thus we have $
W_n(0,t_s,t_w)=W_n(0,|t_s|,|t_w|)$\cite{boundary}. Secondly, we see that the
spectrum $
W_{n-2}(V^{\pm },t_s^{\pm },t_w^{\pm })$ is just uniformly translated from $
W_{n-2}(0,t_s^{\pm },t_w^{\pm })$ by a factor $V^{\pm }$. Thirdly, the
renormalized parameters have the following property:
\begin{equation}
\label{contractionfactor1}
\begin{array}{c}
|t_s^{\pm }|=zt_s\ ,\ |t_w^{\pm }|=zt_w\ ,\ |t_s^0|=\bar zt_s\ ,\
|t_w^0|=\bar zt_w \\
z=\frac{\displaystyle{t_w}}{\displaystyle{2t_s}}\ll 1\ \ \bar z=\frac{%
\displaystyle{t_B^2}}{\displaystyle{t_A^2}}\ll 1
\end{array}
\end{equation}
Combining these properties and using (\ref{niunori}), we deduce the
following new renormalisation scheme:
\begin{equation}
\label{niunori1}W_n(\displaystyle{0,t_s,t_w})\longrightarrow \left|
\begin{array}{l}
-t_s+zW_{n-2}(0,t_s,t_w) \\
\bar zW_{n-3}(0,t_s,t_w) \\
+t_s+zW_{n-2}(0,t_s,t_w)
\end{array}
\right.
\end{equation}
We see that if $t_s$ is sufficiently strong, the spectrum $
W_{n-2}(0,t_s,t_w) $, which is contracted by a factor $z$ and centered
around $\pm t_s$, is not mixed with the spectrum $W_{n-3}(0,t_s,t_w)$, which
is contracted by $\bar z$. More precisely, if we call $\Delta_n$ the
distance between the lowest and highest energy levels of $H_n$, then, this
non--overlapping condition becomes $(z\Delta_{n-2}+\bar z\Delta_{n-3})\le
2t_s$. Under this condition, relation (\ref{niunori1}) gives the following
recurrence scheme between the energy levels $E_n^i(0,t_s,t_w)$, $
E_{n-2}^i(0,t_s,t_w)$ and $E_{n-3}^i(0,t_s,t_w)$:
\begin{equation}
\label{niunori6}
\begin{array}{llr}
E_n^i & =-t_A+zE_{n-2}^i & (i=1,F_{n-2}) \\
E_n^{i+F_{n-2}} & =\bar zE_{n-3}^i & (i=1,F_{n-3}) \\
E_n^{i+F_{n-1}} & =t_A+zE_{n-2}^i & (i=1,F_{n-2})
\end{array}
\end{equation}
Similarly, the associated recurrence for the bandwidths is given by:
\begin{equation}
\label{niunori7}
\begin{array}{ll}
\Delta _n^i & =z\Delta _{n-2}^i \\
\Delta _n^{i+F_{n-2}} & =\bar z\Delta _{n-3}^i \\
\Delta _n^{i+F_{n-1}} & =z\Delta _{n-2}^i
\end{array}
\end{equation}

We now give the quantitative consequences of the last three relations (\ref
{niunori1})(\ref{niunori6})(\ref{niunori7}) upon the spectral properties.
{}From the recurrence scheme (\ref{niunori6}) we see that we can assign a set
of indices $\{+\}(+t_s)$, $\{0\}$ or $\{-\}(-t_s)$ to each individual energy
level, according to the $path$ $of$ $bifurcations$ of that level. Therefore a
typical level $E_n^i$ has $n_{+}$, $n_0$ and $n_{-}$ indices
($+,0,-$); with the constraint $2(\negthinspace n_{+}\negthinspace
+\negthinspace n_{-}\negthinspace )\negthinspace +\negthinspace
3n_0\negthinspace =\negthinspace n(\pm 1)$. As examples, the lowest level is
indexed by $\{---...\},(n_{-}=[n/3],n_{+}=n_0=0)$ and the highest is indexed
by $\{+++...\}$\cite{review}\cite{indexation2}. From this indexation and
relation (\ref{niunori7}), we see that the band associated with a level $
E_n^i(n_{+},n_0,n_{-})$ has a width $\Delta _n^i(p,q)\approx z^p\bar z^q$
with $p\negthinspace =\negthinspace (n_{+}\negthinspace +\negthinspace
n_{-}) $ and $q\negthinspace =\negthinspace n_0$. Consequently, the number
of bands of width $\Delta _n(p,q)$ is given by $N_n(p,q)\negthinspace =
\negthinspace 2^p\left( _p^{p+q}\right) $. These last two results allow us to
calculate the exponents $\alpha $ and $g(\alpha )$ defined in the introduction
($ \Delta _n^i\negthinspace =\negthinspace F_n^{-1/\alpha _i}$ and $N_n(p,q)
\negthinspace =\negthinspace F_n^{g(\alpha )}$). As shown in relation (\ref
{multifrac6}), in the quasiperiodic limit ($n\rightarrow \infty $), these
exponents are function of the parameter $x=p/n$ which varies continuously in
$[0,1/2]$\cite{zheng}.
\begin{equation}
\label{multifrac6}
\begin{array}{ll}
\alpha (x) & =\ln
{\omega }/(x\ln {z/\bar z^{2/3}}+\ln {\bar z^{1/3}}) \\ g(\alpha (x)) &
=(x\ln
{3x/2}-(1+x)\ln {(1+x)^{1/3}} \\  & \ \ +(1-2x)\ln {(1-2x)^{1/3}})/\ln {
\omega }
\end{array}
\end{equation}
{}From this last relation we can obtain two interesting properties of the
spectrum. Firstly, we observe that when $z=\bar z^{2/3}$($t_w/t_s=1/8$), the
exponent $\alpha $ is independent of $x$. As a consequence the spectrum is a
pure fractal with Hausdorff dimension $D_F=\alpha =\ln {\omega ^3}/\ln {\bar
z}$. In contrast, when for example $z>\bar z^{2/3}$, we see that the
exponent $\alpha $ varies between $\alpha _{min}=\ln {\omega ^3}/\ln {\bar z}
$ and $\alpha _{max}=\ln {\omega ^2}/\ln {z}$. As these two exponents
correspond to the bandwidth associated with the central and edge levels
respectively, this property allows us to compare their values with the exact
analytical result of Kohmoto\cite{kohmoto2}\cite{kohmoto3}.From this we
observe that our shrinking factors $z$ and $\bar z$ are just the dominant
terms in a $(t_w/t_s)$ series expansion of the two more exact values $%
z_{ex}=2/(\sqrt{(J-1)^2-4}+J-1)$ and $\bar z_{ex}=1/(\sqrt{1+4(I+1)^2}%
+2(I+1))$ where $I=\frac 14(\frac{t_w}{t_s}-\frac{t_s}{t_w})^2$ and $J=3+
\sqrt{25+16I}$.

So far, we have only analyzed the properties of individual bandwidths and
levels. However it is also possible and very instructive to study integrated
quantities.We start with the spectral measure and a physical quantity
closely related to it. The spectral measure $d\mu _n(E)$ of the hamiltonian $
H_n$ is $d\mu _n(E)=\rho _n(E)dE$ where $\rho _n(E)=\frac
1{F_n}\sum_{i=1}^{F_n}\delta (E-E_n^i)$ is the density of states. From this
definition and relation (\ref{niunori6}) we deduce the following recurrence
\cite{bellstinchcombe}:
\begin{equation}
\label{mesure1}
%\begin{array}{ll}
%d\mu _n(E)= & \omega _n^2d\mu _{n-2}(
%\frac{E+t_s}z)+\omega _n^3d\mu _{n-3}(\frac E{\bar z}) \\  & \ \ +\omega
%_n^2d\mu _{n-2}(\frac{E-t_s}z)
%\end{array}
d\mu _n(E)=\omega _n^2d\mu _{n-2}(\frac{E+t_s}z)+\omega _n^3d\mu _{n-3}(\frac
E{\bar z}) +\omega_n^2d\mu _{n-2}(\frac{E-t_s}z)
\end{equation}
As an example of the application of relation (\ref{mesure1}), we calculate the
large time average return probability defined by $p_n(t)=|\int_{-\infty
}^{+\infty }e^{-iEt}d\mu _n(E)|^2=2\pi \tilde \mu (t)\tilde \mu^{*} (t)$
\cite{geisel}. Using (\ref{mesure1}) we firstly deduce that $\tilde \mu
_n(t)=2\omega _n^2\cos {t_st}\tilde \mu _{n-2}(zt)+\omega _n^3\tilde \mu
_{n-3}(\bar zt)$. Now, in the large time limit, we have on average $\langle
\cos {t_st}\rangle \sim 0$ and $\langle \cos {}^2{t_st}\rangle \sim 1/2$, and
from this we immediately get:
\begin{equation}
\label{mesure2}p_n(t)=2\omega _n^4p_{n-2}(zt)+\omega _n^6p_{n-3}(\bar zt)
\end{equation}
In the limit $(n\rightarrow \infty )$, we see that a fixed point solution of
relation (\ref{mesure2}) is $p^{*}(t)\sim t^{-\gamma }$ where the exponent $%
\gamma $ is determined by $2\omega ^4z^{-\gamma }+\omega ^6\bar z^{-\gamma
}=1$. As we now show, this exponent $\gamma $ is one of the anomalous
dimensions $D_q$ that characterize the MF properties of the spectral
measure. In our case, these non trivial dimensions $D_q$ are defined by the
requirement that the partition function\cite{halsey} $\Gamma _n(q,\tau
=(q-1)D_q)=F_n^{-q}\sum_{i=1}^{F_n}(\Delta _n^i)^{-\tau }$ be stationary in
the limit $n\rightarrow \infty $. Using relation (\ref{niunori7}) we get the
following recurrence for the $\Gamma _n(q,\tau )$\cite{zheng}:
\begin{equation}
\label{multifrac1}\Gamma _n(q,\tau )=2\frac{\displaystyle{\omega _n^{2q}}}{%
\displaystyle{z^\tau }}\Gamma _{n-2}(q,\tau )+\frac{\displaystyle{\omega
_n^{3q}}}{\displaystyle{\bar z^\tau }}\Gamma _{n-3}(q,\tau )
\end{equation}
The stationary constraint then gives a self--consistent equation for the $D_q
$
\begin{equation}
\label{multifrac3}2\omega ^{2q}z^{(1-q)D_q}+\omega ^{3q}\bar z^{(1-q)D_q}=1
\end{equation}
{}From this last relation, we immediately see that the exponent $\gamma $
previously defined is in fact equal to $D_2$. A second consequence of
(\ref{multifrac3}) is that the Hausdorff dimension $D_F=D_0$ is the solution
of $2z^{D_F}+\bar z^{D_F}=1$. To see further the use of relation (\ref
{multifrac3}) and the role of the $D_q$ we calculate two other quantities of
interest: the Lebesgue measure $B_n=\sum_{i=1}^{F_n}\Delta _n^i$;
and the number of bands of width between $\Delta $ and $\Delta+d\Delta $,
$dN_n(\Delta )=\sum_{i=1}^{F_n}\delta (\Delta -\Delta_n^i)d\Delta $.
Using relation (\ref{niunori7}) we can easily deduce a recurrence relation for
each of these quantities:
\begin{equation}
\label{lebesgue}B_n=2zB_{n-2}+\bar zB_{n-3}
\end{equation}
\begin{equation}
\label{bande2}dN_n(\Delta )=2dN_{n-2}(\frac \Delta z)+dN_{n-3}(\frac \Delta
{\bar z})
\end{equation}
The first of these equations was partially guessed in \cite{kohmoto1} for a
model on the Fibonacci chain; a very similar relation was also derived
for the case of the Harper model in\cite{bellstinchcombe}. From equation (
\ref{lebesgue}), we can show that the large $n$ behavior of $B_n$ is $
B_n\simeq B_0F_n^{-\delta }$ where the exponent $\delta $ is related to the
anomalous dimensions by $D_{-\delta }=\frac 1{1+\delta }$. Now, considering (
\ref{bande2}), we see that in the limit $n\rightarrow \infty $, a possible $
invariant$ form is given by $dN^{* }(\Delta )=\Delta ^{-(1+\beta)}d\Delta $
with $\beta =D_F$. In view of relation (\ref{multifrac6}), this simple $
invariant$ solution is quite surprising and indeed its sense is not very
clear.In particular, if instead of $dN_n(\Delta )$
we look at the distribution of bandwidths $W_n(\Delta )=\frac
{dN_n(\Delta)}{F_nd\Delta}$, then the corresponding recurrence relation allows
the two possible invariant distributions $W^{*}(\Delta )=\delta (\Delta )$
and $W^{*}(\Delta )=\Delta^{-1}$, whose form do not coincide with
$W^{*}(\Delta )=\frac{dN^{* }(\Delta )}{d\Delta}=\Delta
^{-(1+D_F)}$\cite{banddistri}

To complete the study of the spectral properties, we also look at the the
statistical properties of the gaps. Roughly speaking a gapwidth is the
distance between two levels. In consequence, we might expect their
distribution to be quite similar to that of the bandwidths. However, there
is an important difference; the number of gaps of a chain $F_n$ is $F_n-1$,
thus, $2(F_{n-2}-1)+F_{n-3}-1=F_n-3<F_n-1$! Looking at figure \ref{fig1} the
last inequality means that if we take only gaps coming from $zW_{n-2}(\pm
t_s,t_s,t_w)$ and $\bar zW_{n-3}(0,t_s,t_w)$, we miss two gaps which are in
fact precisely the biggest. Taking this into account, we see that the number of
gaps of width between $g$ and $g+dg$,
$dN_n(g)=\sum_{i=1}^{F_n}\delta (g-g_n^i)dg$, obeys the following recurrence
($n\ge 3$):
\begin{equation}
\label{gaps1}dN_n(g)=2dN_{n-2}(\frac{\displaystyle{g}}{\displaystyle{z}}
)+dN_{n-3}(\frac{\displaystyle{g}}{\displaystyle{\bar z}})+2\delta (g-g_n^0)dg
\end{equation}
where the last term re--introduces  the two largest gaps of width $g_n^0$ at
each iteration. A further refining of our description requires two other
important remarks. The first is that the initial conditions are $dN_0(g)=0
$, $dN_1(g)=0$ and $dN_2(g)=\delta (g-g_0)dg$. The second is that in the
limit $n\rightarrow \infty $ the width $g_n^0$ tends to a fixed value $g^{*
}=t_A-\frac 12(\bar z\Delta ^{*}+z\Delta ^{*})=t_A\frac{1-\bar
z-2z}{(1-z)}$ where $\Delta ^{*}$ is the width of
the spectrum in this limit. This last remark allows to replace the previous
recurrence for $dN_n(g)$ by the following effective equation which describes
the
infinite size behavior more effectively:
\begin{equation}
\label{gaps2}dN_n(g)\simeq 2dN_{n-2}(\frac{\displaystyle{g}}{\displaystyle{z}}%
)+dN_{n-3}(\frac{\displaystyle{g}}{\displaystyle{\bar z}})+2\delta
(g-g^{* })dg
\end{equation}
As shown in table \ref{table1} the iteration of relation (\ref{gaps2}), with
the previous initial conditions, produces two kinds of gaps. The first kind
corresponds to what we call the $transient$ gaps. For a chain $F_n$, these
gaps have widths of the form $g=z^p\bar z^qg_0$ with ($2p\negthinspace
+\negthinspace 3q\negthinspace =\negthinspace n-2$) and $N_n(p,q)
\negthinspace =\negthinspace 2^p\left( _p^{p+q}\right) $ (last two columns
of table \ref{table1}). These $transient$ gaps are created by iteration of
the initial condition $dN_2(g)=\delta (g-g_0)dg$ and their effective recurrence
does not contain the last term of (\ref{gaps2}). For these reasons their
distribution is strictly similar to that of bandwidths and in particular the
sums of their widths decreases like the spectrum Lebesgue measure $B_n$. In
contrast, the second kind of gaps are those created by the presence of the
last term in relation (\ref{gaps2}) (first two columns of table \ref{table1}
). As can be seen in table \ref{table1}, if a gap of this kind opens for, say a
chain $F_p$, it persists for longer chains; it is $stable$. For a
such a chain $F_n$, these $stable$ gaps have widths of the form $g=z^p\bar
z^qg^{*}$ with $N_n(p,q)\negthinspace =\negthinspace 2^p\left( _p^{p+q}\right)
$ but now $2p+3q$ takes all values between $0$ and $n-3$. Due to this
difference
the distribution of the $stable$ gaps differs from that of the transient
gaps in the following way: in a similar manner to (\ref{bande2}), the
absence of the last term yields an $invariant$ solution to (\ref{gaps2}) of
the form $dN^{*}(g)=g^{-(1+D_F)}dg$. However, similarely to
bandwidths\cite{banddistri}, for the $transient$ gaps, $dN_n(g)$ tends to
$dN^{* }(g)$. In contrast when the last term is present,  that is for
the $stable$ gaps, then, the function $dN_n(g)$ really tends to $dN^{*
}(g)=g^{-(1+D_F)}dg$ over the whole interval $[0,g^{* }]$. Due to this
property,in that case, we can also define a distribution which is of the form $
P^{*}(g)=\frac{dN^{*}(g)}{dg}=g^{-(1+D_F)}$. An additional property of the
$stable$ gaps concerns the value ($G_n$) of the sum of their widths. From (\ref
{gaps2}) we see that $G_n$ obeys a recurrence relation $G_n=2zG_{n-2}+\bar
zG_{n-3}+2g^{*}$, and from this we can deduce that in contrast to $B_n$, $G_n$
does not decrease with $F_n$ but tends to a value $G^{*}$ which is
exactly equal to the spectrum width $\Delta ^{*}$.

In conclusion, we have described as completely as possible the statistical
properties of the energy spectrum of a tight--binding hamiltonian
on the Fibonacci chain. We have compared several of the new predictions with
numerical computations with satisfactory results. As the text has made
clear, our results, qualitatively and quantitatively
complete and unify previous works on similar models. \\
%More generally we think that results like $D_{-\delta}=\frac{1}{1+\delta}$ and
%%$P^{*}(g)=g^{-(1+D_F)}$, are valid for all MF measures which obey a
%%reccurence relation like (\ref{mesure1}) (eg. $\mu_n(x)=\sum_{i}
%%\mu_{n-i}(f_i(x))$ where $f_i(x)$ is a linear function of $x$ so that the
%%different $\mu_{n-i}(f_i(x)))$ do not $overlapp$).
%The comparison of the analytical results describded here, with direct
%%numerical simulations, reveal discrepancies which we can qualitatively
%%explain\cite{fred1}.
% To be complete we
%should mention a similar analytical study can also be done for the phonon
%case by the replacement $E_n^i(\pm )\rightarrow (-\Omega _n^i)^2+t_w+t_s)$, $
%E_n^i(0)\rightarrow (-\Omega _n^i)^2+t_w+t_w)$ and $I\rightarrow \frac{
%(\Omega _n^i)^2}4(\frac 1{t_s}-\frac 1{t_w})^2$, where $\Omega _n^i$ is the
%phonon frequency\cite{fred1}.

$Aknowledgement$: One of us (F.P) thanks Jean Bellissard for very useful
clarifications, and the Lab. Phys. Quantique Universit\'e Paul Sabatier for
its kind hospitality. F.P would also like to thank Aude Lalanne and Th\'eo
Berthin for their encouragement during this work.

%}
\begin{table}
\caption{Gapwidths obtained for the six first iterations of relation
$(15)$. Stable gaps are written in the first column. Transient gaps
are in the third column.}
\label{table1}
\begin{tabular}{l|l|l|l|l|}
{$F_n$} &
 \multicolumn{1}{c|}{$g/g^{*}$}
&\multicolumn{1}{c|}{$N(g)$}
&\multicolumn{1}{c|}{$g/g_0$}
&\multicolumn{1}{c|}{$N(g)$}\\ \hline
2&\#&\#&$1$&$1$\\
3&$1$&$2$&\#&\#\\
5&$1$&$2$&$z$&$2$\\
8&$1 \ z$&$2 \ \ 4$&$\bar{z}$&$1$\\
13&$1 \ z \ \bar{z}$&$2 \ \ 4 \ \ 2$&$z^2 $&$4$\\
21&$1 \ z \ \bar{z} \ z^2$&$2 \ \ 4 \ \ 2 \ \ 8$&$z\bar{z}$&$4$\\
34&$1 \ z \ \bar{z} \ z^2 \ z\bar{z}$&$2 \ \ 4 \ \ 2 \ \ 8 \ \ 8$&$z^3
\ \bar{z}^2$&$8 \ \ 1$\\
55&$1 \ z \ \bar{z} \ z^2 \ z\bar{z} \ z^3 \ \bar{z}^2$&$ 2 \ \ 4 \ \ 2
\ \ 8 \ \ 8 \ \ 16 \ \ 2$&$z^2\bar{z}$&$12$\\
\end{tabular}
\end{table}
%\begin{figure}
%\vglue=8.cm
%\special{psfile=fibo.ps
%angle=90
%vscale=20 hscale=20 hoffset=328 voffset=328}
%\epsfxsize=6.in
%\epsfbox{../fibo.ps}
%\protect \caption{Spec\-trum of ap\-pro\-xi\-mant ha\-mil\-to\-nian
%$(H_n,F_n)$, $n\le8$ de\-duce from re\-la\-tion $(5)$ for
%$(t_s\!=\!1,t_w\!=\!0.5)$. The three initial conditions are,
%$(H_0,F_0\!=\!1,t_i\!=\!t_s)$; $(H_1,F_1\!=\!1,t_i\!=\!t_w)$ and
%$(H_2,F_2\!=\!2,t_{2i}\!=\!t_s,t_{2i+1}\!=\!t_w)$. $z$ and $\bar{z}$ are the
%two shrincking factor. $g_O$ and $g^{*}$, are the initial transient gap and
%%the
%maximum stable gap respectively.} \protect
%\label{fig1}
%\end{figure}
\begin{figure}
\protect \caption{Spec\-trum of the ap\-pro\-xi\-mant ha\-mil\-to\-nian
$(H_n,F_n)$, $n\le8$ de\-duced from re\-la\-tion $(5)$ for
$(t_s\!=\!1,t_w\!=\!0.5)$. The three initial conditions are,
$(H_0,F_0\!=\!1,t_i\!=\!t_s)$; $(H_1,F_1\!=\!1,t_i\!=\!t_w)$ and
$(H_2,F_2\!=\!2,t_{2i}\!=\!t_s,t_{2i+1}\!=\!t_w)$. $z$ and $\bar{z}$ are the
two shrinking factors. $g_O$ and $g^{*}$, are the initial transient gap and the
maximum stable gap respectively.} \protect
\label{fig1}
\epsfxsize=12.cm
%\epsfbox{fibo.ps}
\end{figure}

\begin{thebibliography}{99}
\bibitem{boundary}For finite chain, this is only valid for the Bloch phase
$k=\pi/2$ ($\sim$ band centers).
\bibitem{banddistri}  There are many other problems with distributions
related to the bandwidths and the way we take the limit $n\rightarrow \infty $
. All these problems come from the fact that both the individual bandwidth
and their degeneracy are too much strongly fluctuating variables. Due to
this, even if $dN^{*}(\Delta )\ vs\ \Delta$ is an $invariant$ of (\ref{bande2}
), for any finite chain $F_n$ the $function$ $dN_n(\Delta )$ does not tend to
$dN^{*}(\Delta )$. In fact, only the function $g(\alpha )$ has a meaning.
\end{thebibliography}
\end{document}